\documentclass[aip,graphicx,apl,reprint]{revtex4-1}
\usepackage{graphicx}
\usepackage{txfonts}

\draft 

\begin{document}
\preprint{}

\title{Probing quasiparticle excitations in a hybrid single electron transistor} 
\author{H. S. Knowles}
\affiliation{Low Temperature Laboratory (OVLL), Aalto University, P.O. Box 15100, FI-00076 AALTO, Finland}
\affiliation{Cavendish Laboratory, University of Cambridge, JJ Thomson Avenue, Cambridge CB3 0HE, United Kingdom}
\author{V. F. Maisi} 
\affiliation{Low Temperature Laboratory (OVLL), Aalto University, P.O. Box 15100, FI-00076 AALTO, Finland}
\affiliation{Centre for Metrology and Accreditation (MIKES), P.O. Box 9, FI-02151 Espoo, Finland}\author{J. P. Pekola}
\affiliation{Low Temperature Laboratory (OVLL), Aalto University, P.O. Box 15100, FI-00076 AALTO, Finland}

\date{\today}

\begin{abstract}
We investigate the behavior of quasiparticles in a hybrid electron turnstile with the aim of improving its performance as a metrological current source. The device is used to directly probe the density of quasiparticles and monitor their relaxation into normal metal traps. We compare different trap geometries and reach quasiparticle densities below 3~$\muup$m$^{-3}$ for pumping frequencies of 20 MHz. Our data show that quasiparticles are excited both by the device operation itself and by the electromagnetic environment of the sample. Our observations can be modelled on a quantitative level with a sequential tunneling model and a simple diffusion equation.
\end{abstract}

\maketitle 

Applications of superconductors generally rely on the fact that electronic excitations can be generated only if energy higher or equal to the gap energy $\Delta$ is available. Hence the number of the excitations is ideally exponentially small at low temperatures and the properties specific for superconductors appear. If the number of excitations, typically characterized by the density of quasiparticles, increases, the superconducting features degrade. Such an effect has been studied in several devices such as superconducting qubits~\cite{Martinis2009, Paik2011, corcoles2011, Visser2011}, superconductor-insulator-normal metal-insulator-superconductor (SINIS) microcoolers~\cite{Pekola2000, Rajauria2009-2, Oneil2011,arutyunov2011} and kinetic inductance detectors~\cite{Day2003, Barends2008}. In this letter we focus on the effects of quasiparticles on a hybrid single-electron transistor. We use the Coulomb blockaded transistor for direct and simple probing of the quasiparticle excitation density. The device operates as a charge pump and is a promising candidate for the realization of a metrological current source~\cite{Pekola2008}. We show experimentally that the quasiparticle excitations limit the current quantization. By optimizing the quasiparticle relaxation, however, we estimate that it is possible to reach metrological accuracy.

\begin{figure}[t]
\includegraphics[width=8.5cm]{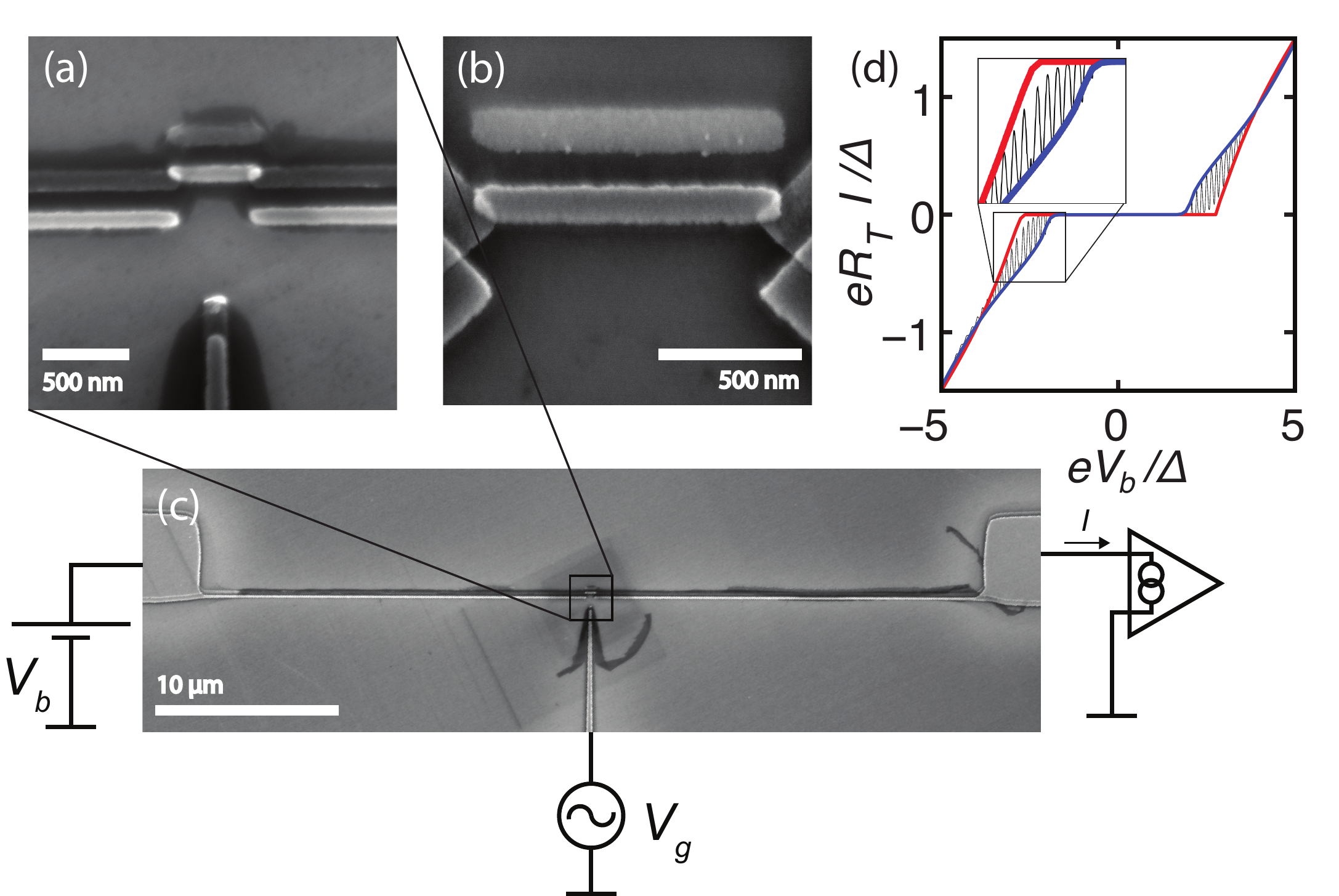}
\caption{(a) Scanning electron micrograph of the $l = 20$ $ \muup$m type A SINIS20 turnstile showing the superconducting Al leads connected to the Cu metal island via AlO$_2$ barrier junctions and the gate electrode used to regulate the potential of the island. (b) shows the type B SINISopen geometry sample where the normal metal traps are within 200 nm of the junctions and the leads open up straight from the junction. In (c) the full length $l$ of the superconducting line of sample SINIS20 between the junction and the trap is visible as well as the quasiparticle traps formed by overlapping Cu and Al shadows. A sketch of the basic measurement circuitry is depicted and (d) shows the IV characteristics of the turnstile.}
\label{sample}
\end{figure}

In order to observe how quasiparticles influence the performance of the turnstile we designed samples (type A) where the quasiparticle relaxation in normal metal traps was purposefully delayed by extending the bare superconducting lines that connect the junctions to the traps. The beginning of this isolated superconducting line can be seen in the scanning electron microscope image of the turnstile in Fig.~\ref{sample}.~(a) where it connects to the normal metal island via oxide junctions that appear as lighter areas. Its extension is visible in (c) all the way through to the wide traps of overlapping superconductor and normal metal separated by an oxide layer. The sample shown in (b) (type B) has wide leads with normal metal traps close to the junctions to enable efficient quasiparticle evacuation. The samples were fabricated with the standard electron-beam lithography and shadow mask technique \cite{PhysRevLett.59.109}. We compare the behaviour of quasiparticles in SINIS turnstile samples with different geometries. The length of the isolated superconducting line (given by the separation of the transistor junction from the trap) was varied between $l = 200$  nm and $l = 20$ $\muup$m. Measurements were performed in a dilution refrigerator at a base temperature of approximately 60 mK.

Figure \ref{sample} (d) shows the current-voltage characteristics of the turnstile sample presented in (a) and (c). The black line shows the current through the turnstile when the bias voltage V$_b$ is swept across the superconducting gap and the gate potential V$_g$ is varied between the gate open ($n_g = 0.5$) and the gate closed ($n_g = 0$) states. Simulations based on sequential tunnelling are fitted to the data in order to extract the parameters of the superconducting gap $\Delta = 216\ \mathrm{\muup eV}$, the charging energy of the island $E_c = 0.74 \Delta$ and the tunnelling resistance of the junction $R_T = 91$ k$\Omega$, specific to each sample (see Table~1 for parameters of all samples measured). These simulations are shown in blue for the gate open and red for the gate closed state. On this coarse level, the heating of the superconducting leads does not make a significant contribution and can be neglected. The overheating of the normal metallic island (thickness 30 nm) is taken into account by considering the electron-phonon coupling with material parameter value of $\Sigma = 2 \cdot 10^9\ \mathrm{WK^{-5}m^{-3}}$ which is consistent with values obtained in previous experiments~\cite{Kafanov2009,Timofeev2009}. In the subgap and pumping experiments we have the opposite situation: the island does not heat since the signals are much smaller but as we look at small differences, the overheating of the superconductor starts to have an effect.

\begin{figure}
\includegraphics[width=8.5cm]{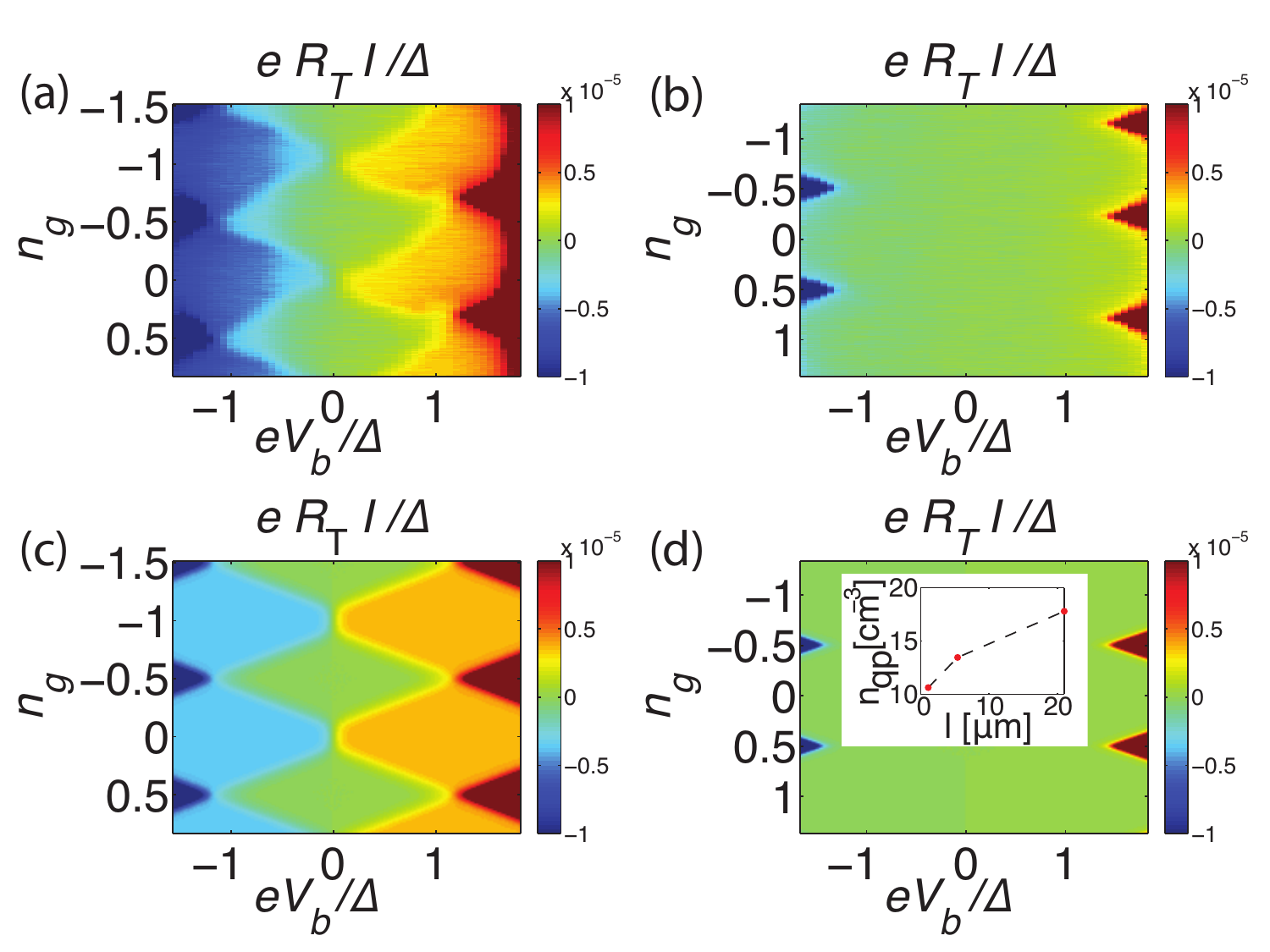}
\caption{(a) and (b) show the normalized current through the oxide trap turnstile of geometry $l = 20$ $\muup$m smoothed over 5 data points as a function of $V_b$ for gate voltages varying between $n_g = 0$ and $n_g = 0.5$. The charge stability measurement plotted in (a) was performed in a sample stage with a single cover; (b) shows measurements of the same sample in an indium sealed two-cover stage. In (c) and (d)  we plot the normalized current computed with a simulation based on sequential tunneling for the situations of (a) and (b) respectively. The inset of (d) shows the quasiparticle density $n_{qp}$ inferred from the temperature reproducing the measured current inside the superconductor gap as a function of the length $l$ of the isolated superconducting line. 
}
\label{DC}
\end{figure}

We now investigate the current of the turnstile more closely for bias voltages within the superconducting gap. These measurements allow the direct probing of the quasiparticle density. This is not possible with voltage biased NIS junctions that probe the excitations more indirectly~\cite{Ullom1998, Pekola2000, Oneil2011, Rajauria2009-2,arutyunov2011} and correspond to the gate open case in these experiments. Figure~\ref{DC}~(a) shows the current as a function of $V_b$ and $V_g$. With no quasiparticles present we would expect zero current for low biases. However, we observe a current pattern periodic in $V_g$ within the gap with currents rising up to 10 fA. We find that we can reproduce the measured current pattern with a simulation shown in panel (c) using a high superconductor temperature $T_S$ that gives rise to a quasiparticle population above the gap. We compute the average current through the left junction via $I = e \sum_{n}(\Gamma_{LI} (n) - \Gamma_{IL} (n)) P(n,t)$ where $\Gamma_{LI(IL)}$ is sequential tunnelling rate to (from) the island through the left junction and $P(n,t)$ is the probability of the system being in the charge state $n$ of the island. The temperatures used to fit these data are $T_S = 205$ mK and the normal metal temperature $T_N = 92$~mK. 
The current-voltage characteristics are surprising: At degeneracy (half integer $n_g$) no net current flows, whereas in Coulomb blockade (integer $n_g$) we obtain a finite current. The simulations give us an insight into the on-going processes: at degeneracy the hot quasiparticle excitations lying at high energies are able to tunnel in both directions equally, hence there is no net current. In Coulomb blockade, the tunneling of a quasiparticle excitation is followed by a fast relaxation to the lowest lying charge state. The relaxation always happens in the forward direction given by $V_b$ and leads to a net current through the device. These features are a strong indication of having quasiparticle excitations as the source of the sub-gap current in the device.

We measured the same sample in two different sample stages. Instead of having an enclosed stage with one metallic cover as in the measurement displayed in Fig. \ref{DC} (a) we also used an indium sealed double hermetic metallic sample stage.
In the latter case, the radio frequancy (RF) line had an additional 20 cm thermocoax cable to enhance the sample shielding. This was not included in the wiring of the single cover stage. However, as the RF line is not electrically connected
to the turnstile, we do not expect this to influence the direct heat conduction to the sample. The dc wiring of the two setups was similar, made of approximately two meters of thermocoax cable.
The two sample stages are thermalized to the cryostat base temperature in an identical way and the sole purpose of this sealing is to create an RF shield to the sample. The result is shown in panel (b). In this case no sub-gap current can be resolved. This behavior was fitted with temperatures of $T_S \leq 167$~mK and $T_N = 72$~mK which we present in Fig. 2 (d). The inset in Fig.~\ref{DC}~(d) displays the quasiparticle density $n_{qp}$ inferred from the relation $n_{qp} = 2D(E_F) \int_0^\infty n_S(E) e^{-\beta E} \,dE = \sqrt{2\pi}\ D(E_F) \Delta \sqrt{k_BT/\Delta}\ e^{-\Delta /k_B T}$ which is valid at low temperatures $k_BT_S << \Delta$ for three different samples with varying distance to the trap and measured with the poorly filtered sample stage. We use $D(E_F) = 1.45 \times 10^{47} m^{-3}J^{-1}$ as the normal state density of states at the Fermi energy\cite{Kittel} . We observe a monotonous increase in $n_{qp}$ with increasing distance of the trap from the junction. We deduce that the presence of environmentally excited quasiparticles is determined by both the trap relaxation rate and the diffusion rate through the superconducting line.

\begin{figure}
\includegraphics[width=8.5cm]{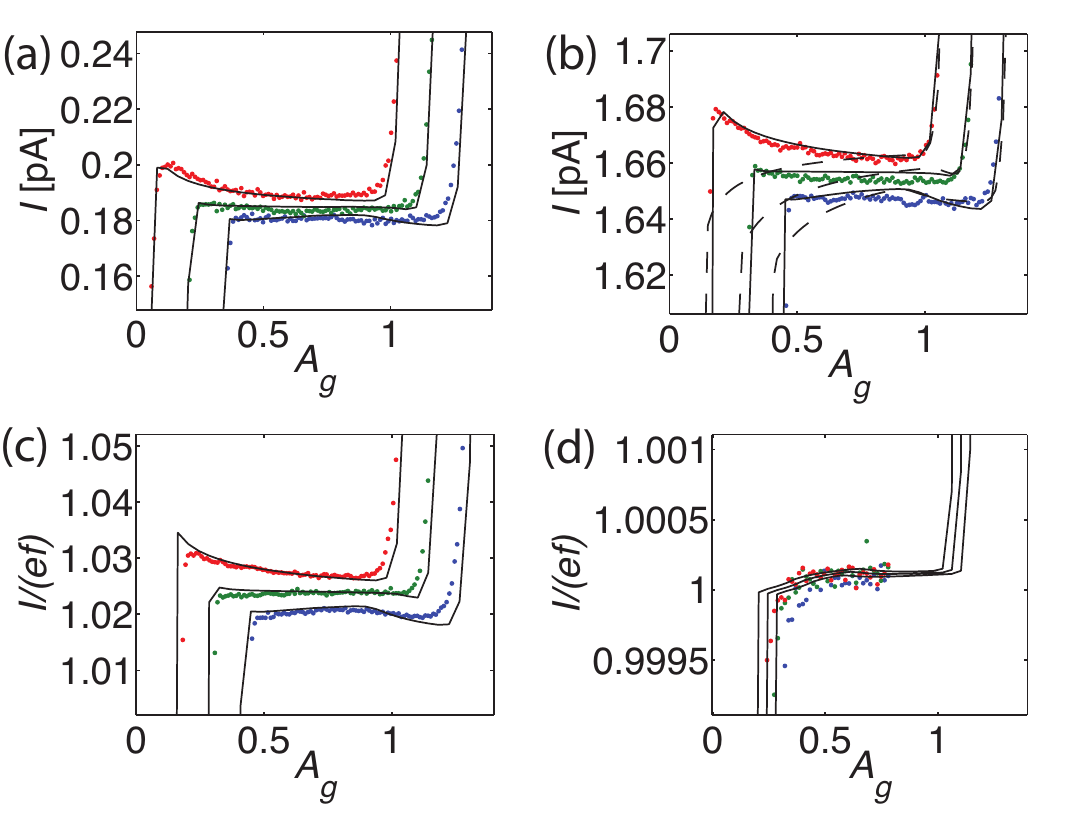}
\caption{Current plateaus under pumping. Current through the turnstile as a function of the gate modulation amplitude measured on the type A SINIS20 sample at (a) 1 MHz, (b) 10 MHz and (c) 20 MHz sinusoidal $V_g$ modulation frequency and on the type B sample SINISopen in (d) at 20 MHz. The pump was operated at bias voltage values of $eV_b = 0.8 \Delta $ (blue dots), $eV_b = 1.2\Delta$ (green dots) and $eV_b = 1.6 \Delta$ (red dots). Fits from simulations including two-electron Andreev processes are displayed as solid black lines, dashed black lines are simulations only including single electron processes.}
\label{pumping}
\end{figure}

Next we turn to the dynamic case of the turnstile operation. In addition to environmental excitation, quasiparticles are now injected to the superconducting leads once in every pump cycle. The pumping frequency thus allows us to control the injected power and the number of quasiparticles. Figure \ref{pumping} shows current plateaus measured on type A SINIS20 sample. Three bias voltages were chosen around the optimum operation voltage of $eV_b = \Delta$ \cite{AverinPekola2008} ranging from 0.8$\Delta /e$ to 1.6$\Delta /e$. Two main effects were observed: a slight overshoot for the highest bias voltage at the beginning of the plateau and a spreading of the plateau value for different $V_b$. Simulations including single electron and two-electron Andreev processes were used to model the process. The peak at the beginning of the plateau can be reproduced by allowing for an Andreev current with a conduction channel area of 30 nm$^2$. This value is taken from previous experiments with similar samples~\cite{Aref2011, Maisi2011}. The spreading of the plateaus in $V_b$ is caused by quasiparticle excitations and follows the expected behaviour with respect to variations in geometry and pumping operation as described below.

The spread increases with the pumping frequency $f$ from $\Delta I = 8$ fA for 1 MHz (panel (a) of Fig.~\ref{pumping}) to $\Delta I = 15$ fA for 10 MHz (panel (b)). The injected power $P_{inj}$ increases with $f$ and we model this increase in mean number of quasiparticles in the superconductor by raising its temperature $T_S$. The bias-dependence of current on the plateau is not influenced by Andreev currents and we can thus fit it simply by ascribing it to the quasiparticle number. The black lines correspond to simulations with quasiparticle densities of $n_{qp} = 41.1\ \muup$m$^{-3}$ at 1 MHz and $n_{qp} = 82.6\ \muup$m$^{-3}$ at 10 MHz.

To demonstrate that the quantization can be improved by enhancing the quasiparticle relaxation we measured the pumping plateaus of a sample with broader leads, see Fig.~\ref{sample}~(b). Panels (c) and (d) in Fig.~\ref{pumping} show the comparison of poor and good quasiparticle trapping respectively. The current quantization improves by two orders of magnitude when using leads with enhanced quasiparticle relaxation. 
 
To model the relaxation of the quasiparticle excitations we consider their diffusion in a thin superconducting line. The heat diffusion equation is 
\begin{equation}
\label{eq:diff}
\nabla \cdot \left(- \kappa_S \nabla T \right) = -p_{\mathrm{trap}},
\end{equation} 
where the thermal conductivity of the superconductor is $\kappa_S = \frac{6}{\pi^2} \left( \frac{\Delta}{k_BT}\right) ^2 e^{-\Delta/k_BT} L_0T/\rho_n$ with the Lorentz number $L_0$ and the normal state resistivity $\rho_n$ \cite{Bardeen1959,Abrikosov1988}. The heat is removed from the superconductor by a normal metallic trap to which it is tunnel coupled. The trap removes the heat $p_{\mathrm{trap}}~=~\frac{2 \sigma_T}{e^2 d} \int_0^\infty E n_S(E) (f_N(E) - f_S(E))  \,dE$ per unit area where $\sigma_T$ is the conductance of the trap per unit area and $d = 22$ nm is the thickness of the superconducting film. The power $P_{inj}$ injected into the line during turnstile operation sets a boundary condition to the beginning of the line: $P_{inj} = A \cdot (-\kappa_S \nabla T)$, where $A = wd$ is the cross-sectional area of the lead with width $w$. We rewrite Eq.~(\ref{eq:diff}) in terms of quasiparticle density $n_{qp}$ by considering only the strong exponential dependencies on $T$ to obtain
\begin{equation}
\label{eq:nqpDiff}
\nabla^2 n_{qp} = \lambda^{-2}(n_{qp}-n_{qp0}),
\end{equation} 
where $\lambda^2 = \frac{\sqrt{2}d}{\sqrt{\pi}\rho_n \sigma_T} \left(\frac{k_BT}{\Delta}\right)^{1/2}$ and $n_{qp0}$ is the quasiparticle density of the superconductor when it is fully thermalised to the normal metal. Next we solve Eq.~(\ref{eq:nqpDiff}) for three different sections of the bias line for type A samples. In the first part we treat the bare aluminium line of length $l$ and constant width $w_1$ with no quasiparticle trapping ($p_{trap} = 0$). The next part deals with the widening of the lead from $w_1$ to $w_2$ where we also neglect quasiparticle trapping since its contribution in this section of the lead is small for our samples. Then, in the last part, the lead continues with a constant cross section of $w_2$ and is in contact with the quasiparticle trap: $p_{trap} \neq 0$. As a result, we obtain the quasiparticle density at the junction to be
\begin{equation}
\label{eq:nqpDensity}
n_{qp} = \frac{\sqrt{\pi}e^2 D(E_F) \rho_n}{\sqrt{2\Delta k_B T}} \frac{P_{inj}}{wd} \left(l + w_1 \log \left( \frac{w_1}{w_2}\right) + \lambda \frac{w_1}{w_2}  \right).
\end{equation} 
The first term is the diffusion in the bare aluminium wire, the second term is the spreading to the wider line and the last term arises from the relaxation to the trap. Similarly, we can solve the diffusion for the type B sample with opening bias lines. Here we assume that the line starts at radius $r_0$ and the injected power is distributed evenly to all directions. We take $r_0 = 70\ \mathrm{nm}$ so that the area the power is injected into, $\frac{\pi}{2} r_0 d = (50\ \mathrm{nm)^2}$, matches the junction area of the sample. The quasipartice density at the junction is then
\begin{equation}
\label{eq:nqpDensityOpening}
n_{qp} = \frac{\sqrt{\pi}e^2 D(E_F) \rho_n}{\sqrt{2\Delta k_B T}}\frac{P_{inj}}{\theta r_0 d} \frac{K_0(r_0/\lambda)}{K_1(r_0/\lambda)},
\end{equation} 
where $K_n$ is the modified Bessel function of second kind and $\theta$ the opening angle of the line.

\begin{figure}
\includegraphics[width=8.5cm]{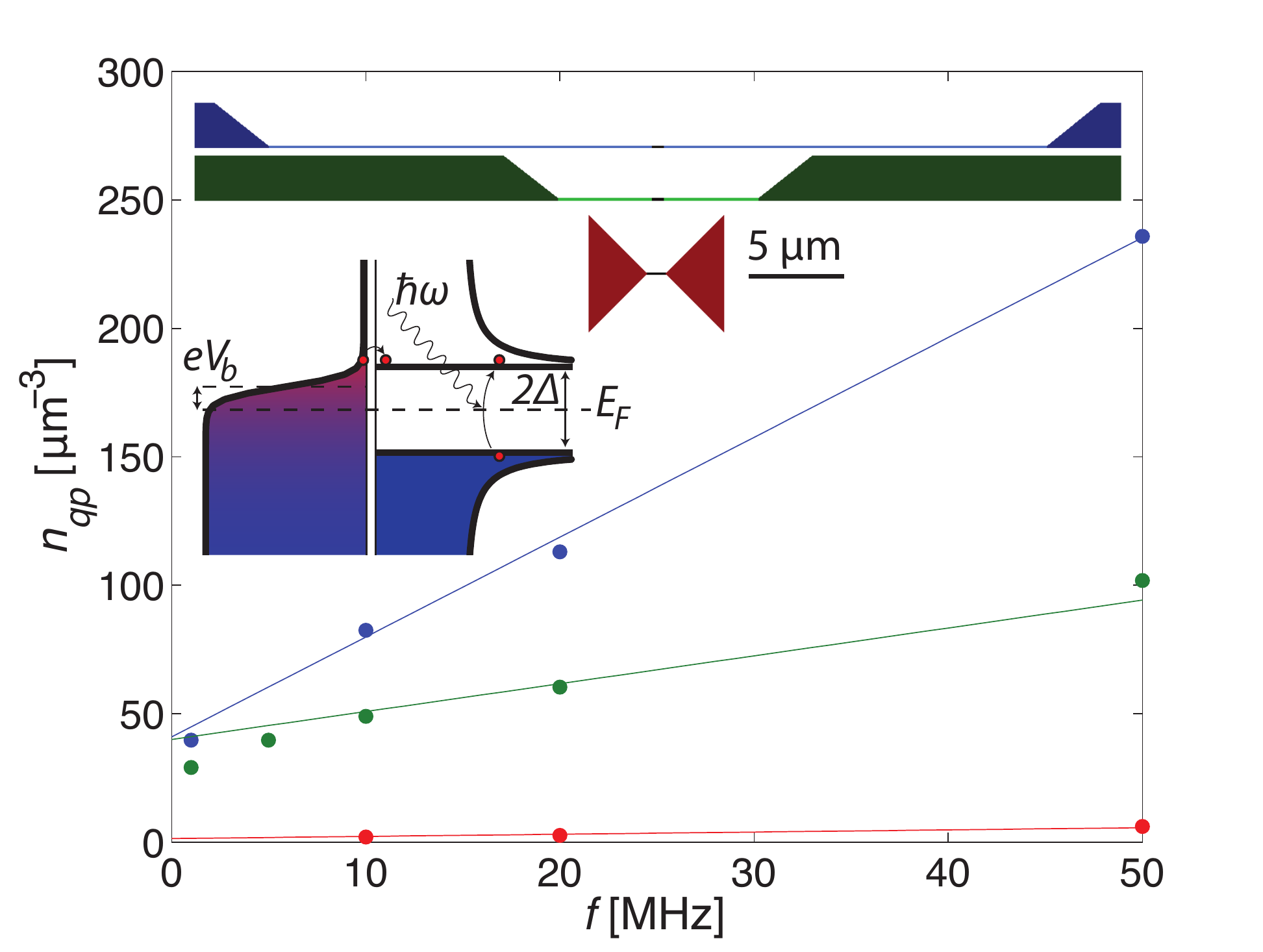}
\caption{Quasiparticle density as a function of the driving gate voltage $f$. Fits to measurements including first and second order tunnelling processes are displayed as dots and quasiparticle densities derived from the diffusion model are displayed as solid lines. We compare two type A samples with delayed relaxation, SINIS20 (blue) and SINIS5 (green) and one type B sample SINISopen (red). Diagrams of the sample geometries are displayed in the upper part of the figure, dark colours representing quasiparticle traps, lighter areas the isolated superconducting lines and the small black bulks the normal metal islands. Below, a diagram shows the two pathways to quasiparticle excitation, during pumping via the normal metal island and through interactions with the electromagnetic environment.}
\label{nqpVSf}
\end{figure}

We now compare the quasiparticle relaxation in different sample geometries. From fits to measurements similar to those shown in Fig. \ref{pumping} we extract $n_{qp}$ as a function of $f$. These values are displayed as dots in Fig. \ref{nqpVSf} for the SINIS20 (blue), the SINIS5 (green) and the SINISopen samples (red). Using the diffusion model described above we calculate $n_{qp}$ as a function of $f$ corresponding to an injection power on the plateau given by $P_{inj} = ef\Delta$ (solid lines). The injected heat calculated from the simulations deviated less than 10 \% from this power even at the highest frequencies measured. We used $\rho_n = 31\ \mathrm{n\Omega m}$ as the normal state resistivity for all samples. Individual sample parameters used in the simulations are listed in Table 1.
The measured densities clearly show a linear behavior with increasing $f$ as predicted by the model (see Eqs. (3) and (4)). As indicated by the slopes for increasing length $l$ of the isolated line, the further the quasiparticle trap lies from the oxide junction and the thinner the connecting line is, the slower the relaxation process is and the more the superconductor is heated. For the samples SINIS20 and SINIS5 a finite quasiparticle density is observed even in the absence of injected power. This density is of the same order of magnitude as the leakage currents shown in Fig.~\ref{DC} and it points towards interactions with electromagnetic radiation from the environment. The SINISopen sample clearly shows the most efficient quasiparticle relaxation with only a few quasiparticles per $\muup$m$^{-3}$ at 50 MHz driving frequency.

\begin{table}[ht]
\setlength{\tabcolsep}{2pt}
\caption{Sample parameters}
\centering
\begin{tabular}{ l c c c c c c}
\hline
Sample & $E_c/\Delta$ & $R_T$ [k$\Omega_1$]& $l\ [\muup$m] & $w_1$ [nm] & $w_2 [\muup m]$ & $\sigma_T [(\Omega m^2)^{-1}]$
\\ [0.5ex]
\hline
SINIS20	&	0.74	&	91	&	21	&	120  &	5 	& 	2.3$\times10^9$	\\ 
SINIS5	&	0.7	&	81 	&	5.4	&	120	&	5 	& 	2.7$\times10^9$\\
SINIS1	&	0.7	&	85	&	1.1	&	160	&	5 	& 	9.5$\times10^8$\\
SINISopen&	1.2	&	170	&	-	&	-	& 	-	 & 	1.0$\times10^9$\\[1ex]
\hline
\end{tabular}
\label{table}
\end{table}

We have investigated the process of quasiparticle relaxation in hybrid SINIS turnstiles for various geometries. Both direct current and gate-driven pumping measurements can be well understood and simulated using models based on first and second order tunnelling processes. We were also able to model the relaxation of the excitations by a simple diffusion model. We find that the main sources for quasiparticles in the single electron transistor are those injected above the superconducting band gap during turnstile operation and those excited by radiation from a hot environment. In the best structure studied in this Letter we reached an accuracy $\delta I /I$ of the order of $1 \cdot 10^{-4}$. We estimate that accuracy better than $10^{-6}$ at $50\ \mathrm{MHz}$ would be obtained with the following improvements: The aluminum should be made an order of magnitude thicker and the trap ten times more transparent. Alternatively, the bias leads can be extended to the third dimension. The resistance of the junctions should be increased by a factor of four and charging energy to $E_c>  2 \Delta$. Finally, the density of environmentally activated quasiparticles needs to be reduced to a level $n_{qp} \ll 0.1\ \mathrm{\mu m^{-3}}$ which has been demonstrated experimentally in  a recent work~\cite{saira2012}. With these realistic modifications we expect to reach metrological accuracy in turnstile operation.

\begin{acknowledgments}
We thank A. Kemppinen and O.-P. Saira for useful discussions and M. Meschke for technical assistance. The work has been supported partially by the National Doctoral Programme in Nanoscience (NGS-NANO) and the European Community's FP7 Programme under Grant Agreements No. 228464 (MICROKELVIN, Capacities Specific Programme) and No. 218783 (SCOPE).
\end{acknowledgments}

\end{document}